\documentclass[%
reprint,
 showpacs,
 amsmath,amssymb,
]{revtex4-1}

\usepackage{graphicx}
\usepackage{dcolumn}
\usepackage{bm}

\usepackage{CJK}

\usepackage{color}
\usepackage{amssymb}
\usepackage{amsfonts}

\usepackage[colorlinks,urlcolor=blue,linkcolor=blue,citecolor=blue,anchorcolor=blue]{hyperref}

\begin{document}

\preprint{APS/123-QED}

\title{Non-Gaussian Nature and Entanglement of Spontaneous Parametric Nondegenerate Triple-Photon Generation}


\author{Da Zhang$^{1,3}$}
\author{Yin Cai$^1$}
\email{caiyin@xjtu.edu.cn}
\author{Zhan Zheng$^2$}
\author{David Barral$^3$}
\author{Yanpeng Zhang$^1$}
\email{ypzhang@mail.xjtu.edu.cn}
\author{Min Xiao$^{4,5}$}
\author{Kamel Bencheikh$^3$}
\email{kamel.bencheikh@c2n.upsaclay.fr}

\affiliation{%
$^1$\mbox{Key Laboratory for Physical Electronics and Devices of the Ministry of Education \& Shaanxi Key Lab of Information }
\mbox{Photonic Technique, School of Electronic and Information Engineering, Xi'an Jiaotong University, Xi'an 710049, China} \\
$^2$\mbox{Zhejiang Provincial Key Laboratory of Quantum Precision Measurement, College of Science, Zhejiang University} \\
\mbox{of Technology, Hangzhou 310023, China}  \\
$^3$\mbox{Centre de Nanosciences et de Nanotechnologies CNRS/Universit$\acute{e}$ Paris-Saclay, 91460 Marcoussis, France} \\
$^4$\mbox{Department of Physics, University of Arkansas, Fayetteville, Arkansas 72701, USA}
$^5$\mbox{National Laboratory of Solid State Microstructures and School of Physics, Nanjing University, Nanjing 210093, China}
}%

\date{\today}

\begin{abstract}
\noindent
How to prepare deterministically non-Gaussian entangled states is a fundamental question for continuous-variable quantum information technology.
Here, we theoretically demonstrate through numerical methods that the triple-photon state generated by three-photon spontaneous parametric down-conversion is a pure super-Gaussian resource of non-Gaussian entanglement.
Interestingly, the degree of entanglement between the modes of the triple-photon state is stronger than that corresponding to the two-mode squeezed vacuum state produced by a quadratic Hamiltonian with the same parameters.
Furthermore, we propose a model to prepare two-mode non-Gaussian entangled states with tunable non-Gaussianity based on quadrature projection measurements.
\end{abstract}

\maketitle

Early development of continuous-variable quantum information technologies (CV-QIT) mainly focused on Gaussian states and Gaussian operations \cite{vanloock.rmp.77.513.2005,weedrook.rmp.84.621.2012}, achieving remarkable accomplishments \cite{furasawa.science.282.706.1998,Grangier.nature.421.6920.2003,furusawa.nature.431.430.2004}.
In CV-QIT, information is encoded in the amplitude and phase quadratures of optical fields, and retrieved by homodyne detection.
Non-classical sources in CV-QIT are generally single-mode squeezed and two-mode entangled states.
They are Gaussian states because the associated quadratures exhibit normal distribution.
The combination of Gaussian states and Gaussian operations provides a complete framework for many CV-QIT protocols \cite{braunstein.prl.80.869.1998,lam.prl.81.5668.1998,braunstein.pra.61.042302.2000,ralph.pra.66.042321.2002,Ban_1999,cerf.pra.63.052311.2001,scarani.rmp.81.1301.2009}.
However, there has been growing awareness of some important inherent limitations of this Gaussian framework, such as quantum distillation \cite{masahide.np.4.178.2010}, which is an essential protocol for long-distance quantum communication, especially quantum key distribution.
A prominent no-go theorem states that Gaussian operations can not distill Gaussian states \cite{eieret.prl.89.137903.2002}.
It has been demonstrated that non-Gaussian (nG) sources and operations are necessary to achieve universal CV quantum computation \cite{l1oyd.prl.82.1784.1999,bartlett.prl.88.097904.2002,nielsen.prl.97.110501.2006,eisert.pra.82.042336.2010,eisert.pra.85.062318.2012}.

Since the available resources for CV-QIT are by nature mostly Gaussian, one must rely on degaussification via quantum nonlinear processes to produce nG resources.
Ourjoumtsev \emph{et al.} used photon subtraction as nG operation to prepare and increase entanglement between two Gaussian states \cite{ourjoumtsev.prl.98.030502.2007,ourjoumtsev.np.5.189.2009}.
So far, photon subtraction and addition have been employed to prepare various optical nG states, such as superposition of coherent states \cite{ourjoumtsev.science.312.83.2006,sychev.np.11.379.2017,furusawa.prl.121.143602.2018,nicola.prl.124.033604.2020}, hybrid entanglement \cite{jeong.np.8.564.2014,morin.np.8.570.2014}, and multimode nG states \cite{ra.np.2019non}.
However, one of the main advantages of CV-QIT is determinism, but the probabilistic degaussification protocol losses this key feature.

A promising quantum technology that can deterministically generate nG states is the triple-photon generation (TPG) process, in which a pump photon $\omega_P$ is down-converted simultaneously into a triplet ($\omega_A$, $\omega_B$ and $\omega_C$) with phase-matching conditions $\omega_P=\omega_A+\omega_B+\omega_C$ (energy conservation) and $\vec{k}_P=\vec{k}_A+\vec{k}_B+\vec{k}_C$ (momentum conservation).
Douady \emph{et al.} \cite{douady.23.2794.ol.2004} firstly demonstrated the existence of phase-matched TPG in the $\mathrm{KTiOPO_4}$ crystal in the regime of double seeding.
Recently, the outstanding work of observing spontaneous TPG in a superconducting parametric cavity was reported by Sandbo Chang \emph{et al.} \cite{chang.prx.10.011011.2020}.
In a fully degenerate configuration, the Wigner function of the triple-photon state (TPS) exhibits negativities \cite{felbinger.prl.80.492.1998,kamel.crp.8.206.2007,chang.prx.10.011011.2020}, a clear signature of nG statistics, which is impossible to achieve by second-order down-conversion (SODC) and linear optics.
Nonlinear steering has been proposed and verified based on a partially degenerate configuration \cite{lam.prl.114.100403.2015}. 
Bencheikh \emph{et al.} revealed the relationship between Gaussian entanglement and both number of seeded modes and intensity in the nondegenerate case \cite{kamel.prl.120.043601.2018}.
\begin{figure*}[htpb]
\centering
  \includegraphics[width=15.5cm]{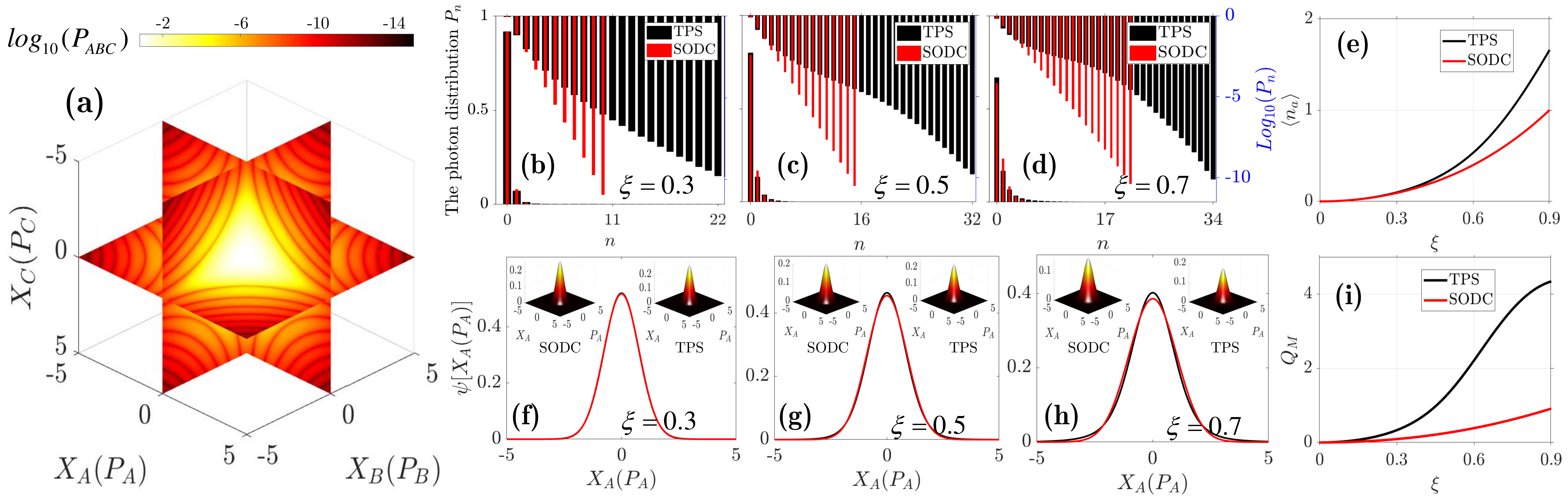}
  \caption{(a) Quadrature joint probability distribution of modes $A$, $B$, and $C$ for interaction strength $\xi=\kappa t\alpha_p=0.3$. The photon number distribution of mode $A$ in linear and log scales for (b) $\xi=0.3$, (c) $\xi=0.5$, (d) $\xi=0.7$, where modes $B$, $C$, and $P$ are traced out. Black and red histograms represent TPS and SODC, respectively, as in the others following curves. (f), (g) and (h) are respectively the marginal distribution of the photon number states represented by (b), (c) and (d). The insets on the left and right in (f), (g) and (h) are respectively the Wigner functions of mode $A$ of TPS and SODC. The mean photon number (e) and Mandel Q parameter (i) of mode $A$.}
  \label{fig1}
\end{figure*}
However, a deeper insight into the physics of nondegenerate TPS in the spontaneous parametric regime is still missing, and whether such states are nG or exhibit entanglement is not clear.

In this letter, we theoretically demonstrate through numerical methods that TPS is a pure super-Gaussian resource of nG entanglement.
We show in particular that the degree of entanglement of TPS is higher than the two-mode squeezed vacuum state produced by the SODC with the same interaction parameters.
We also reveal how a two-mode nG entangled state with tunable non-Gaussianity is prepared by performing simple quadrature projection measurements, and how the entanglement and nG level of the generated states depends on the measurement of quadratures.

We start our analysis by considering the interaction Hamiltonian describing the nondegenerate TPG process
\begin{align}\label{eq1}
\hat{H}&=i\hbar\kappa^{(3)}(\hat{a}^{\dag}\hat{b}^{\dag}\hat{c}^{\dag}\hat{p}-\hat{a}\hat{b}\hat{c}\hat{p}^{\dag}),
\end{align}
where $\kappa^{(3)}$ is the third-order coupling constant that describes the strength of the nonlinear interaction.
The annihilation operators $\hat{a}$, $\hat{b}$, $\hat{c}$ and $\hat{p}$ describe respectively the three down-converted modes and the pump mode.
Using this Hamiltonian, the Schr$\mathrm{\ddot{o}}$dinger equation is solved to deduce the final state of system at time $t$ when considering that the initial state is vacuum for the triplets and a coherent mode $\alpha_p$ for the pump.
The evolution equation can be numerically solved by the Monte Carlo method \cite{molmer.prl.68.580.1993,molmer.josab.10.524.1993,qutip1,qutip2}.
Figure \ref{fig1}(a) shows the quadrature joint probability distribution corresponding to the modes $A$, $B$, and $C$.
The distribution, which was initially spherical in the 3D plot -- modes $A$, $B$, and $C$ are initially vacuum states -- is now very complex exhibiting star shapes and interferences, indicating quantum correlations between them.

We will now discuss the quantum properties of a single-mode of TPS.
For comparison, we also present the photon number distribution of the twin modes generated by SODC and described by the Hamiltonian $H_S=i\hbar\kappa^{(2)}(\hat{a}^{\dag}\hat{b}^{\dag}\hat{p}-\hat{a}\hat{b}\hat{p}^{\dag})$, where $\kappa^{(2)}$ is the second-order nonlinear interaction strength.
To facilitate the comparison of the physical properties of TPS and SODC, we assume $\kappa^{(2)}=\kappa^{(3)}=\kappa$.
Figure \ref{fig1}(b), \ref{fig1}(c) and \ref{fig1}(d) show the photon number distribution of mode $A$ in linear and log scales for interaction strengths $\xi=0.3$, 0.5 and 0.7, where $\xi=\kappa t\alpha_p$.
These processes could be obtained by considering the third-order nonlinear material in a high-finesse cavity or an optimized superconducting parametric cavity.
These photon number distributions of TPS (in black) are quite different from the exponential decay of SODC (in red), the Poisson distribution of coherent states or the Bose-Einstein distribution of thermal states \cite{kok.quantuminformationprocessing.2010,agarwal2012quantum}.
With the increasing of the interaction time, the shape of photon number distribution of TPS gradually goes from an ¡®arc¡¯ to an '$\backsim$' shape in the log scale, whereas it is always linear for SODC.
This nonlinear photon number distribution has implications on the marginal distributions (black) represented in Figs. \ref{fig1}(f), \ref{fig1}(g) and \ref{fig1}(h) along with the Gaussian marginal distribution (red) of SODC.
As the interaction strength increases, the marginal distribution of TPS becomes nG, exhibiting a higher peak, narrower shoulders and longer tails.
The TPS displays a super-Gaussian distribution in the quadratures, demonstrating thus the nG nature of nondegenerate TPS.
Another interesting feature is that although the proportion of vacuum state in the TPS is relatively high, its mean photon number is still lager than that of SODC due to the long-tailed distribution, as shown in Fig. \ref{fig1}(e).
Figure \ref{fig1}(i) shows the Mandel parameter $Q_M$ of TPS, which is defined as $Q_M=(\langle\Delta^2 n\rangle/\langle n\rangle)-1$ \cite{agarwal2012quantum}.
The $Q_M$ of TPS is always greater than zero, indicating the super-Poisson distribution character of the photon number fluctuations.
Besides, the noise fluctuation of the TPS is higher than that of SODC under the same mean number of photons.

After studying the single-mode properties of the TPS, we focus on its multimode nG and entanglement features.
As shown in Figs. \ref{fig1}(f), \ref{fig1}(g) and \ref{fig1}(h), it is difficult to qualitatively grasp the super-Gaussian nature of the TPS from the Wigner function.
This is in contrast with photon subtraction or addition, where the negativity of the Wigner function at the origin of the phase space is a signature of non-Gaussinity.
Thus, a measure of non-Gaussianity based on a different model is necessary here.
We employ the quantum relative entropy (QRE) to quantify the nG nature of the TPS.
The QRE between a given quantum state $\varrho$ and a reference Gaussian state $\gamma$ is defined as follows \cite{marco.pra.78.6.2008}
\begin{figure}[htpb]
\centering
  \includegraphics[width=\columnwidth]{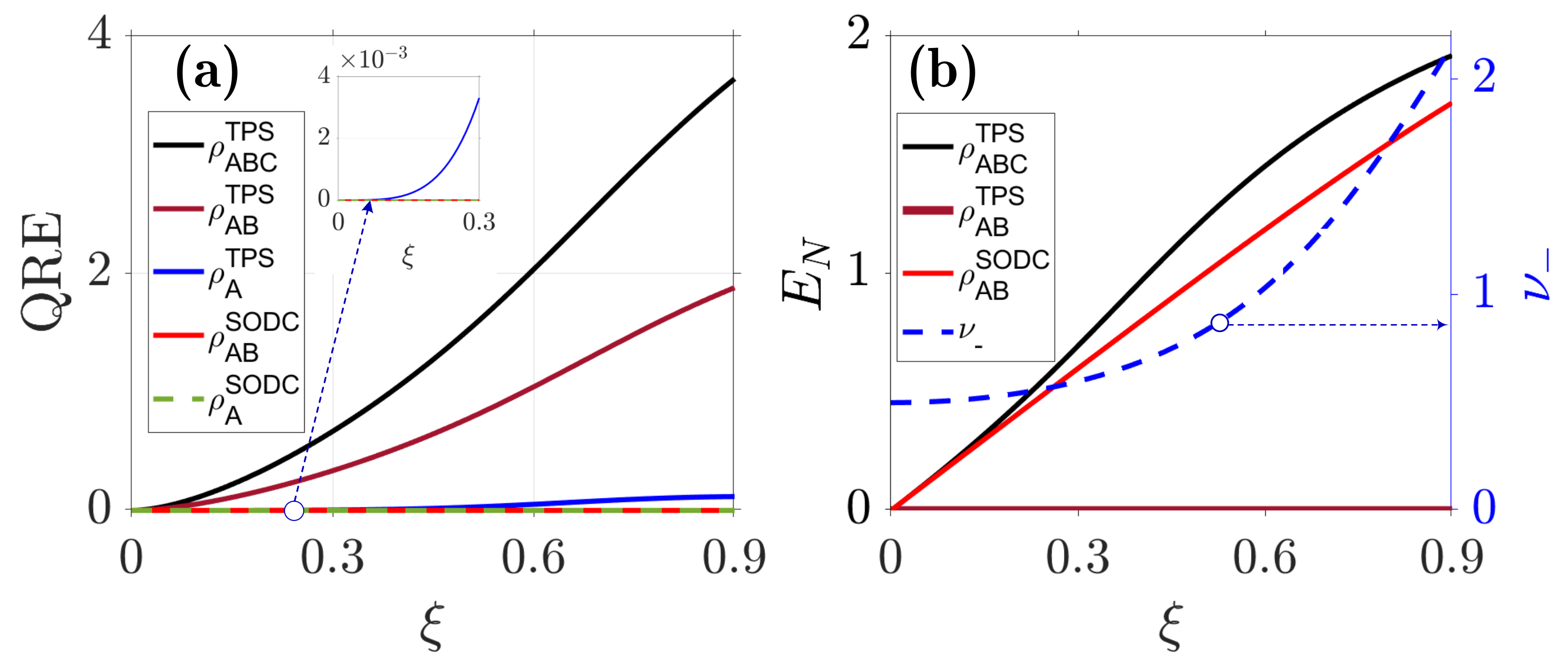}
  \caption{(a) QRE and (b) entanglement of states generated by TPG and SODC, respectively. $\nu_-$ is the smallest symplectic eigenvalue of the partially transposed covariance matrix of the TPS.}
  \label{fig2}
\end{figure}
\begin{align}\label{eq2}
\delta[\varrho]=E_{vN}(\gamma)-E_{vN}(\varrho),
\end{align}
where $E_{vN}(\varrho)=\textrm{Tr}[\varrho\ln\varrho]$ is the von Neumann entropy (vNE), and $\gamma$ has the same covariance matrix and the first moments as $\varrho$.
The vNE of a Gaussian state with covariance matrix $\gamma$ and symplectic eigenvalue $\nu_j$ is
\begin{align}\label{eq3}
E_{vN}(\gamma)=\sum_{j=1}^n f(\nu_j),
\end{align}
where $f(x)=(x+\frac{1}{2})\ln(x+\frac{1}{2}) - (x-\frac{1}{2})\ln(x-\frac{1}{2})$.

Let us first consider the SODC process.
The vNE (entanglement entropy) of the two-mode squeezed vacuum is \cite{kok.quantuminformationprocessing.2010}
\begin{equation}\label{eq4}
E_{vN}^S=\cosh^2(\xi)\ln\cosh^2(\xi)-\sinh^2(\xi)\ln\sinh^2(\xi).
\end{equation}
Figure \ref{fig2}(a) shows in red the QRE [$\delta(\varrho^{SODC}_{AB})=E^S_{vN}(\gamma_{AB})-E_{vN}^S$] of SODC [See the Supplemental Material for detail], where $E^S_{vN}(\gamma_{AB})$ represents the vNE of a reference two-mode Gaussian state.
Not surprisingly, it's equal to zero for all interaction time since $\gamma_{AB} = \varrho^{SODC}_{AB}$.
Any single-mode element of SODC is in a thermal state, and the vNE is $E_{vN}^S(\varrho_{A/B})=E_{vN}^S$.
Similarly, the QRE of single mode [$\delta(\varrho^{SODC}_{A})]$ is equal to zero, as shown in Fig. \ref{fig2}(a) in dashed green line.
Thus, for every Gaussian state independently of the number of modes, the QRE is zero.
\begin{figure*}[htpb]
\centering
  \includegraphics[width=14cm]{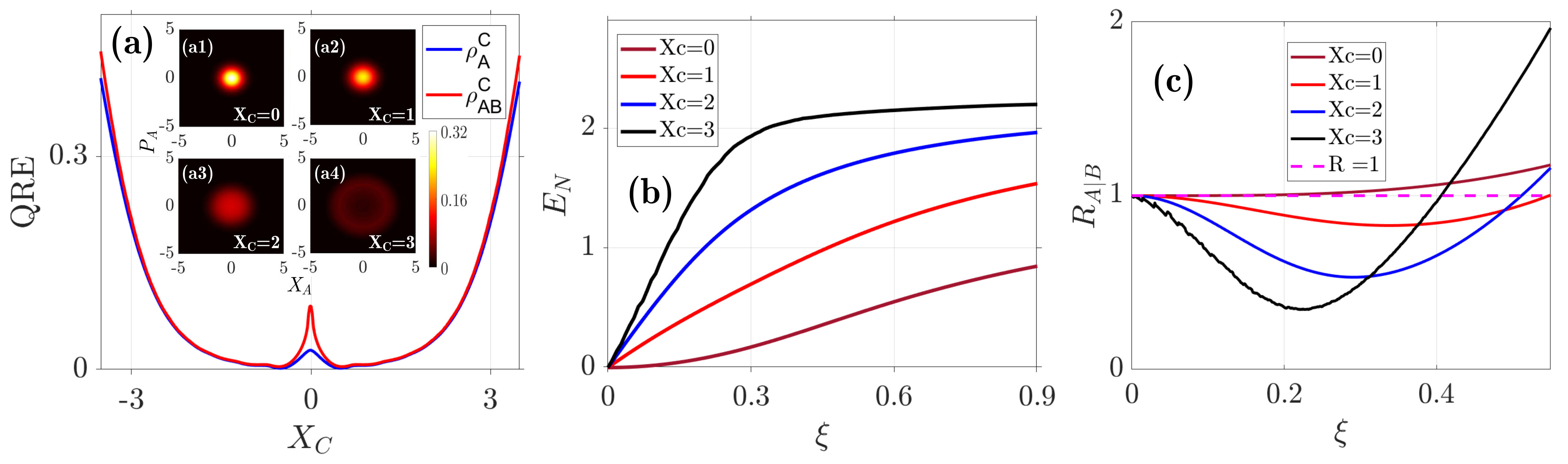}
  \caption{(a) The QRE of conditional two-mode $\rho^C_{AB}$ (red line) and $\rho^C_{A}$ (blue line) versus $X_C$ when $\xi=0.3$. The insets show the Wigner functions of projected mode $A$ when $X_C$=0 (a1), 1 (a2), 2 (a3) and 3 (a4) (Mode $B$ is traced out), respectively. (b) The logarithmic negativity is greater than 0 for all projection measurements indicating entanglement between modes $A$ and $B$. (c) The quadratures displays EPR steering when $X_C\geq1$, and are not EPR steerable when $X_C=0$. The jitter in (b) and (c) is caused by a lower probability of projection measurement when $\xi$ is small.}
  \label{fig3}
\end{figure*}

When the interaction strength is small, the TPS can be approximated as $|\psi_T\rangle\approx(|000\rangle+\xi|111\rangle+\xi^2/2|222\rangle)_{ABC}$.
The entanglement entropy of the TPS is still equal to the vNE of each subsystem.
Due to the symmetry of the system, we obtain $E^T_{vN}(\rho_{A/B/C}) =E^T_{vN}(\rho_{AB/BC/AC})$.
Thus, the QRE mainly depends on the vNE of the reference Gaussian state.
For the TPS represented by $|\psi_T\rangle$, the covariance matrix related to the reference Gaussian state is $\gamma_{ABC}=$Diag$(\lambda,\cdots,\lambda)$, where $\lambda=3\xi^2+5\xi^4/4$.
The vNE of the reference Gaussian state satisfy the relationship $E^T_{vN}(\gamma_{ABC})=3E^T_{vN}(\gamma_{AB/BC/AC})/2=3E^T_{vN}(\gamma_{A/B/C})$.
Therefore, we can deduce a mathematical relation of the QRE between the TPS and its subsystems, given by:
\begin{align}\label{eq5}
\delta^T(\rho_{ABC})-\delta^T(\rho_{AB})=\delta^T(\rho_{AB})-\delta^T(\rho_{A}).
\end{align}
Figure \ref{fig2}(a) displays the evolution of QRE as a function of the interaction strength.
We see that the QRE of TPS increases monotonically throughout the interaction time.
For its subsystems $\rho_{AB}$ and $\rho_{A}$, their QRE still exists as shown in Fig. \ref{fig2}(a) by the brown and blue lines, respectively.
The different results in Fig. \ref{fig2}(a) are calculated numerically without any assumption on the evolved state, validating the relation of Eq. (\ref{eq5}).

Next, we show that the TPS is not only nG, but also entangled.
As an entanglement witness we use the logarithmic negativity defined as $E_N=\ln||\varrho^{T_A}||_1$ \cite{vidal.pra.65.032314.2002}, where $\varrho^{T_A}$ is the partial transpose over mode $A$ and $||\cdot||_1$ is the trace norm.
Obviously, the modes $A$, $B$ and $C$ are entangled and are typically the GHZ state.
This is shown in the logarithmic negativity plots in Fig. \ref{fig2}(b) (black).
On the other hand, the minimum symplectic eigenvalue $\nu_-$ of the partially transposed covariance matrix of TPS is greater than or equal to $1/2$ [$\hat{a}=(\hat{X}_A+i\hat{P}_A)/\sqrt{2}$] for all interaction time [dashed line in Fig. \ref{fig2}(b)], which means that there is no Gaussian entanglement among the modes $A$, $B$ and $C$ \cite{simon.prl.84.2726.2000}.
This is consistent with the experimental results \cite{chang.prx.10.011011.2020}.
The reason for this is that the quantum correlations of the TPS are purely nG, not detected by Gaussian-type inseparability criteria.
Higher-order cumulants need to be calculated for nG statistics.
Remarkably, for the same interaction strength, although the TPS contains more vacuum noise, its degree of entanglement is higher than that of the SODC due to the long-tailed distribution.
Another feature is that when one mode is traced out, the remaining two modes are not entangled, although they are nG, which is also a distinctive signature of GHZ state.
The evolution of the entanglement of the TPS versus the losses is discussed in the Supplemental Material \cite{sm.prl.2020}.

Nowadays, photon addition and subtraction are the most effective ways to prepare two-mode nG states, as it has been implemented in many experiments and used as a common method for quantum distillation.
However, to the best of our knowledge, the nG properties produced by these methods are not readily tunable.
In our proposal, the projection induced by the measurement of a quadrature allows the nG properties to be tuned continuously.
Figure \ref{fig3}(a) shows the dynamic evolution of the QRE of the conditional two-mode ($\varrho^{C}_{AB}$) and arbitrary single-mode ($\varrho^{C}_{A/B}$) versus the projection measurement $X_C$ of mode $C$.
For the outcome $X_C = 0$ in the homodyne detection, the probability distribution is the largest, $\delta[\varrho^{C}_{AB}]$ is great than zero.
In the vicinity of $X_C=0$, as $|X_C|$ increases, the photon number distribution of $\varrho^{C}_{AB}$ gradually tends to decay exponentially, making $\delta[\varrho^{C}_{AB}]$ decrease.
This is a gaussification process.
As $|X_C|$ continues to increase, $\varrho^{C}_{AB}$ starts to deviate from the exponential decay distribution, thus the QRE increases significantly, corresponding to the degaussification process.
The QRE of mode $A$ shows an evolution law similar to that of $\delta[\varrho^{C}_{AB}]$ if mode $B$ is traced out.
In the insets (a1), (a2), (a3), and (a4) of Fig. \ref{fig3}(a), we plot the Wigner functions of mode $A$ for values of $X_C=0, 1, 2$ and 3, respectively.
The nG signature exhibited by the Wigner functions becomes more and more obvious [From (a2) to (a4)], albeit no negativity appears.

Besides, $\varrho^{C}_{AB}$ is inseparable and describes entanglement between modes $A$ and $B$ as shown in Fig. \ref{fig3}(b) representing the logarithmic negativity $E_N$ as a function of the interaction strength and for different $X_C$ outcomes.
With fixed interaction strength, $E_N(\varrho^{C}_{AB})$ increases monotonically with the increase of $X_C$.
However, compared to the TPS, $E_N(\varrho^{C}_{AB})$ is reduced when $X_C=0$.
This is because the projection measurement $X_C=0$ will increase the vacuum noise.
For the $X_C\geq1$, $E_N(\varrho^{C}_{AB})$ is significantly increased compared to the TPS.

For comparison, we studied the linear steering of $\varrho^{C}_{AB}$ \cite{armstrong.np.11.2015}, defined as $R_{A|B}=\Delta(X_A+g_{B,X}X_B)\Delta(P_A+g_{B,P}P_B)<1$, where $g_{B,X}$ and $g_{B,P}$ are optimized real numbers.
The quadrature mode $A$ is steerable by mode $B$ if $R_{A|B}< 1$.
In Fig. \ref{fig3}(c), we see that the quadratures of mode $A$ is steerable by $B$ within a short interaction time when the projection measurement  outcomes are $X_C =1, 2, $ and $3$.
Interestingly, we find that the $R_{A|B}$ is always greater than one under $X_C=0$, that is, there is only nG entanglement between modes $A$ and $B$.
Considering that the TPS contains only nG entanglement, the homodyne detection $X_C=1, 2$ and 3 introduce Gaussian component to $\varrho^{C}_{AB}$, resulting in the coexistence of nG and Gaussian entanglement.

\begin{figure}[htpb]
\centering
  \includegraphics[width=\columnwidth]{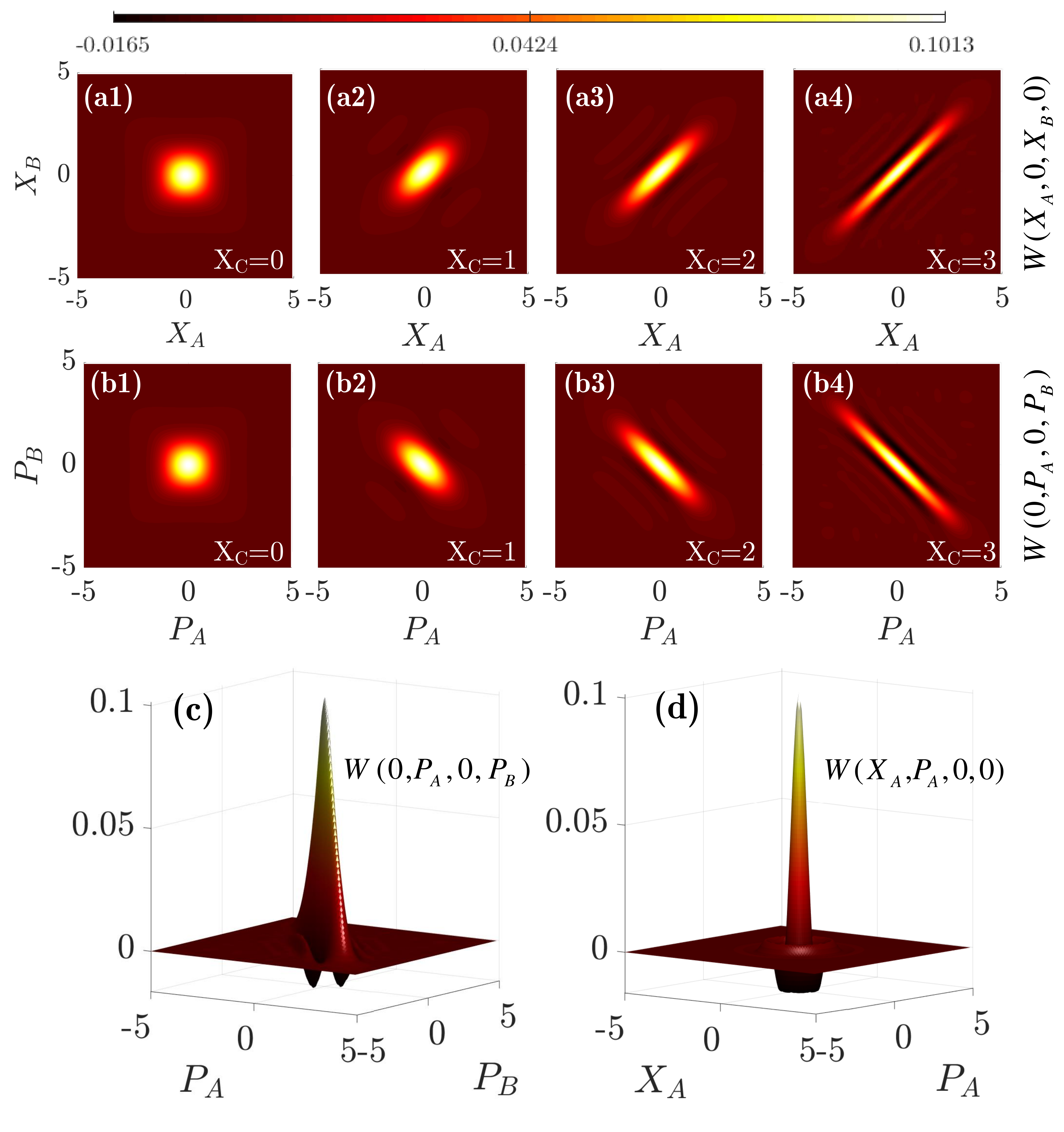}
  \caption{Wigner functions of conditional two-mode. (a1)-(a4) are the Wigner functions $W(X_A,0,X_B,0)$ for values of $X_C=0, 1, 2$ and 3, respectively. (b1)-(b4) are same as (a1)-(a4) but $W(0,P_A,0,P_B)$. (c) and (d) are 3D plots of $W(0,P_A,0,P_B)$ and $W(X_A,P_A,0,0)$ when $X_C=3$, respectively.}
  \label{fig4}
\end{figure}
To visualize the dynamic evolution of nG and entanglement characteristics of $\varrho^{C}_{AB}$ with the projection measurement, we plot the Wigner function of $\varrho^{C}_{AB}$ for the values of $X_C = 0, 1, 2$ and $3$ in Fig. \ref{fig4}.
For an outcome on $X_C = 0$, $W(X_A,0,X_B,0)$ [See Fig. \ref{fig4}(a1)] and $W(0,P_A,0,P_B)$ [See Fig. \ref{fig4}(b1)] reveal a quadruple symmetry (super-Gaussian), and that the amplitude and phase quadratures do not exhibit any correlation, which is also confirmed by the steering in Fig. \ref{fig3}(c).
For the values of $X_C = 1, 2$, and 3, $W(X_A,0,X_B,0)$ [From Fig. \ref{fig4}(a2) to \ref{fig4}(a4)] shows that the amplitude quadratures are correlated while the phase quadratures are anti-correlated in $W(0,P_A,0,P_B)$ [From Fig. \ref{fig4}(b2) to \ref{fig4}(b4)].
Importantly, the correlation gradually increases with the increase of $X_C$.
Finally, we also show 3D plots of $W(0,P_A,0,P_B)$ and $W(X_A,P_A,0,0)$ for an outcome on $X_C=3$ in Fig. \ref{fig4}(c) and \ref{fig4}(d).
The Wigner function show clearly negative values which is a real evidence of the nG nature of the generated states.

In summary, we have theoretically revealed that the triple-photon state produced by triple-photon generation is a pure super-Gaussian entangled state, robust to optical losses \cite{sm.prl.2020}.
Subsequently, we have proposed to use projection measurements to prepare two-mode non-Gaussian entangled states with tunable non-Gaussianity.
Our results extend the boundaries of quantum optics and have potential applications in quantum information processing and quantum computing.
\section{Acknowledgement}
The authors gratefully acknowledge Ariel Levenson for helpful discussions.
This work was supported by the National Key Research and Development Program of China (2017YFA0303700, 2018YFA0307500), National Natural Science Foundation of China (61975159, 61605154, 11604256, 11804267, 11904279), and by the Agence Nationale de la Recherche through Project TRIQUI (No. ANR 17-CE24-0041).

\begin{thebibliography}{45}%
\makeatletter
\providecommand \@ifxundefined [1]{%
 \@ifx{#1\undefined}
}%
\providecommand \@ifnum [1]{%
 \ifnum #1\expandafter \@firstoftwo
 \else \expandafter \@secondoftwo
 \fi
}%
\providecommand \@ifx [1]{%
 \ifx #1\expandafter \@firstoftwo
 \else \expandafter \@secondoftwo
 \fi
}%
\providecommand \natexlab [1]{#1}%
\providecommand \enquote  [1]{``#1''}%
\providecommand \bibnamefont  [1]{#1}%
\providecommand \bibfnamefont [1]{#1}%
\providecommand \citenamefont [1]{#1}%
\providecommand \href@noop [0]{\@secondoftwo}%
\providecommand \href [0]{\begingroup \@sanitize@url \@href}%
\providecommand \@href[1]{\@@startlink{#1}\@@href}%
\providecommand \@@href[1]{\endgroup#1\@@endlink}%
\providecommand \@sanitize@url [0]{\catcode `\\12\catcode `\$12\catcode
  `\&12\catcode `\#12\catcode `\^12\catcode `\_12\catcode `\%12\relax}%
\providecommand \@@startlink[1]{}%
\providecommand \@@endlink[0]{}%
\providecommand \url  [0]{\begingroup\@sanitize@url \@url }%
\providecommand \@url [1]{\endgroup\@href {#1}{\urlprefix }}%
\providecommand \urlprefix  [0]{URL }%
\providecommand \Eprint [0]{\href }%
\providecommand \doibase [0]{http://dx.doi.org/}%
\providecommand \selectlanguage [0]{\@gobble}%
\providecommand \bibinfo  [0]{\@secondoftwo}%
\providecommand \bibfield  [0]{\@secondoftwo}%
\providecommand \translation [1]{[#1]}%
\providecommand \BibitemOpen [0]{}%
\providecommand \bibitemStop [0]{}%
\providecommand \bibitemNoStop [0]{.\EOS\space}%
\providecommand \EOS [0]{\spacefactor3000\relax}%
\providecommand \BibitemShut  [1]{\csname bibitem#1\endcsname}%
\let\auto@bib@innerbib\@empty
\bibitem [{\citenamefont {Braunstein}\ and\ \citenamefont {van
  Loock}(2005)}]{vanloock.rmp.77.513.2005}%
  \BibitemOpen
  \bibfield  {author} {\bibinfo {author} {\bibfnamefont {S.~L.}\ \bibnamefont
  {Braunstein}}\ and\ \bibinfo {author} {\bibfnamefont {P.}~\bibnamefont {van
  Loock}},\ }\href {\doibase 10.1103/RevModPhys.77.513} {\bibfield  {journal}
  {\bibinfo  {journal} {Rev. Mod. Phys.}\ }\textbf {\bibinfo {volume} {77}},\
  \bibinfo {pages} {513} (\bibinfo {year} {2005})}\BibitemShut {NoStop}%
\bibitem [{\citenamefont {Weedbrook}\ \emph {et~al.}(2012)\citenamefont
  {Weedbrook}, \citenamefont {Pirandola}, \citenamefont {Garc\'{\i}a-Patr\'on},
  \citenamefont {Cerf}, \citenamefont {Ralph}, \citenamefont {Shapiro},\ and\
  \citenamefont {Lloyd}}]{weedrook.rmp.84.621.2012}%
  \BibitemOpen
  \bibfield  {author} {\bibinfo {author} {\bibfnamefont {C.}~\bibnamefont
  {Weedbrook}}, \bibinfo {author} {\bibfnamefont {S.}~\bibnamefont
  {Pirandola}}, \bibinfo {author} {\bibfnamefont {R.}~\bibnamefont
  {Garc\'{\i}a-Patr\'on}}, \bibinfo {author} {\bibfnamefont {N.~J.}\
  \bibnamefont {Cerf}}, \bibinfo {author} {\bibfnamefont {T.~C.}\ \bibnamefont
  {Ralph}}, \bibinfo {author} {\bibfnamefont {J.~H.}\ \bibnamefont {Shapiro}},
  \ and\ \bibinfo {author} {\bibfnamefont {S.}~\bibnamefont {Lloyd}},\ }\href
  {\doibase 10.1103/RevModPhys.84.621} {\bibfield  {journal} {\bibinfo
  {journal} {Rev. Mod. Phys.}\ }\textbf {\bibinfo {volume} {84}},\ \bibinfo
  {pages} {621} (\bibinfo {year} {2012})}\BibitemShut {NoStop}%
\bibitem [{\citenamefont {Furusawa}\ \emph {et~al.}(1998)\citenamefont
  {Furusawa}, \citenamefont {S{\o}rensen}, \citenamefont {Braunstein},
  \citenamefont {Fuchs}, \citenamefont {Kimble},\ and\ \citenamefont
  {Polzik}}]{furasawa.science.282.706.1998}%
  \BibitemOpen
  \bibfield  {author} {\bibinfo {author} {\bibfnamefont {A.}~\bibnamefont
  {Furusawa}}, \bibinfo {author} {\bibfnamefont {J.~L.}\ \bibnamefont
  {S{\o}rensen}}, \bibinfo {author} {\bibfnamefont {S.~L.}\ \bibnamefont
  {Braunstein}}, \bibinfo {author} {\bibfnamefont {C.~A.}\ \bibnamefont
  {Fuchs}}, \bibinfo {author} {\bibfnamefont {H.~J.}\ \bibnamefont {Kimble}}, \
  and\ \bibinfo {author} {\bibfnamefont {E.~S.}\ \bibnamefont {Polzik}},\
  }\href {\doibase 10.1126/science.282.5389.706} {\bibfield  {journal}
  {\bibinfo  {journal} {Science}\ }\textbf {\bibinfo {volume} {282}},\ \bibinfo
  {pages} {706} (\bibinfo {year} {1998})}\BibitemShut {NoStop}%
\bibitem [{\citenamefont {Grosshans}\ \emph {et~al.}(2003)\citenamefont
  {Grosshans}, \citenamefont {Van~Assche}, \citenamefont {Wenger},
  \citenamefont {Brouri}, \citenamefont {Cerf},\ and\ \citenamefont
  {Grangier}}]{Grangier.nature.421.6920.2003}%
  \BibitemOpen
  \bibfield  {author} {\bibinfo {author} {\bibfnamefont {F.}~\bibnamefont
  {Grosshans}}, \bibinfo {author} {\bibfnamefont {G.}~\bibnamefont
  {Van~Assche}}, \bibinfo {author} {\bibfnamefont {J.}~\bibnamefont {Wenger}},
  \bibinfo {author} {\bibfnamefont {R.}~\bibnamefont {Brouri}}, \bibinfo
  {author} {\bibfnamefont {N.~J.}\ \bibnamefont {Cerf}}, \ and\ \bibinfo
  {author} {\bibfnamefont {P.}~\bibnamefont {Grangier}},\ }\href
  {https://www.nature.com/articles/nature01289} {\bibfield  {journal} {\bibinfo
   {journal} {Nature}\ }\textbf {\bibinfo {volume} {421}},\ \bibinfo {pages}
  {238} (\bibinfo {year} {2003})}\BibitemShut {NoStop}%
\bibitem [{\citenamefont {Yonezawa}\ \emph {et~al.}(2004)\citenamefont
  {Yonezawa}, \citenamefont {Aoki},\ and\ \citenamefont
  {Furusawa}}]{furusawa.nature.431.430.2004}%
  \BibitemOpen
  \bibfield  {author} {\bibinfo {author} {\bibfnamefont {H.}~\bibnamefont
  {Yonezawa}}, \bibinfo {author} {\bibfnamefont {T.}~\bibnamefont {Aoki}}, \
  and\ \bibinfo {author} {\bibfnamefont {A.}~\bibnamefont {Furusawa}},\ }\href
  {https://www.nature.com/articles/nature02858} {\bibfield  {journal} {\bibinfo
   {journal} {Nature}\ }\textbf {\bibinfo {volume} {431}},\ \bibinfo {pages}
  {430} (\bibinfo {year} {2004})}\BibitemShut {NoStop}%
\bibitem [{\citenamefont {Braunstein}\ and\ \citenamefont
  {Kimble}(1998)}]{braunstein.prl.80.869.1998}%
  \BibitemOpen
  \bibfield  {author} {\bibinfo {author} {\bibfnamefont {S.~L.}\ \bibnamefont
  {Braunstein}}\ and\ \bibinfo {author} {\bibfnamefont {H.~J.}\ \bibnamefont
  {Kimble}},\ }\href {\doibase 10.1103/PhysRevLett.80.869} {\bibfield
  {journal} {\bibinfo  {journal} {Phys. Rev. Lett.}\ }\textbf {\bibinfo
  {volume} {80}},\ \bibinfo {pages} {869} (\bibinfo {year} {1998})}\BibitemShut
  {NoStop}%
\bibitem [{\citenamefont {Ralph}\ and\ \citenamefont
  {Lam}(1998)}]{lam.prl.81.5668.1998}%
  \BibitemOpen
  \bibfield  {author} {\bibinfo {author} {\bibfnamefont {T.~C.}\ \bibnamefont
  {Ralph}}\ and\ \bibinfo {author} {\bibfnamefont {P.~K.}\ \bibnamefont
  {Lam}},\ }\href {\doibase 10.1103/PhysRevLett.81.5668} {\bibfield  {journal}
  {\bibinfo  {journal} {Phys. Rev. Lett.}\ }\textbf {\bibinfo {volume} {81}},\
  \bibinfo {pages} {5668} (\bibinfo {year} {1998})}\BibitemShut {NoStop}%
\bibitem [{\citenamefont {Braunstein}\ and\ \citenamefont
  {Kimble}(2000)}]{braunstein.pra.61.042302.2000}%
  \BibitemOpen
  \bibfield  {author} {\bibinfo {author} {\bibfnamefont {S.~L.}\ \bibnamefont
  {Braunstein}}\ and\ \bibinfo {author} {\bibfnamefont {H.~J.}\ \bibnamefont
  {Kimble}},\ }\href {\doibase 10.1103/PhysRevA.61.042302} {\bibfield
  {journal} {\bibinfo  {journal} {Phys. Rev. A}\ }\textbf {\bibinfo {volume}
  {61}},\ \bibinfo {pages} {042302} (\bibinfo {year} {2000})}\BibitemShut
  {NoStop}%
\bibitem [{\citenamefont {Ralph}\ and\ \citenamefont
  {Huntington}(2002)}]{ralph.pra.66.042321.2002}%
  \BibitemOpen
  \bibfield  {author} {\bibinfo {author} {\bibfnamefont {T.~C.}\ \bibnamefont
  {Ralph}}\ and\ \bibinfo {author} {\bibfnamefont {E.~H.}\ \bibnamefont
  {Huntington}},\ }\href {\doibase 10.1103/PhysRevA.66.042321} {\bibfield
  {journal} {\bibinfo  {journal} {Phys. Rev. A}\ }\textbf {\bibinfo {volume}
  {66}},\ \bibinfo {pages} {042321} (\bibinfo {year} {2002})}\BibitemShut
  {NoStop}%
\bibitem [{\citenamefont {Ban}(1999)}]{Ban_1999}%
  \BibitemOpen
  \bibfield  {author} {\bibinfo {author} {\bibfnamefont {M.}~\bibnamefont
  {Ban}},\ }\href {\doibase 10.1088/1464-4266/1/6/101} {\bibfield  {journal}
  {\bibinfo  {journal} {J. Opt. B-Quantum S. O}\ }\textbf {\bibinfo {volume}
  {1}},\ \bibinfo {pages} {L9} (\bibinfo {year} {1999})}\BibitemShut {NoStop}%
\bibitem [{\citenamefont {Cerf}\ \emph {et~al.}(2001)\citenamefont {Cerf},
  \citenamefont {L\'evy},\ and\ \citenamefont
  {Assche}}]{cerf.pra.63.052311.2001}%
  \BibitemOpen
  \bibfield  {author} {\bibinfo {author} {\bibfnamefont {N.~J.}\ \bibnamefont
  {Cerf}}, \bibinfo {author} {\bibfnamefont {M.}~\bibnamefont {L\'evy}}, \ and\
  \bibinfo {author} {\bibfnamefont {G.~V.}\ \bibnamefont {Assche}},\ }\href
  {\doibase 10.1103/PhysRevA.63.052311} {\bibfield  {journal} {\bibinfo
  {journal} {Phys. Rev. A}\ }\textbf {\bibinfo {volume} {63}},\ \bibinfo
  {pages} {052311} (\bibinfo {year} {2001})}\BibitemShut {NoStop}%
\bibitem [{\citenamefont {Scarani}\ \emph {et~al.}(2009)\citenamefont
  {Scarani}, \citenamefont {Bechmann-Pasquinucci}, \citenamefont {Cerf},
  \citenamefont {Du\ifmmode~\check{s}\else \v{s}\fi{}ek}, \citenamefont
  {L\"utkenhaus},\ and\ \citenamefont {Peev}}]{scarani.rmp.81.1301.2009}%
  \BibitemOpen
  \bibfield  {author} {\bibinfo {author} {\bibfnamefont {V.}~\bibnamefont
  {Scarani}}, \bibinfo {author} {\bibfnamefont {H.}~\bibnamefont
  {Bechmann-Pasquinucci}}, \bibinfo {author} {\bibfnamefont {N.~J.}\
  \bibnamefont {Cerf}}, \bibinfo {author} {\bibfnamefont {M.}~\bibnamefont
  {Du\ifmmode~\check{s}\else \v{s}\fi{}ek}}, \bibinfo {author} {\bibfnamefont
  {N.}~\bibnamefont {L\"utkenhaus}}, \ and\ \bibinfo {author} {\bibfnamefont
  {M.}~\bibnamefont {Peev}},\ }\href {\doibase 10.1103/RevModPhys.81.1301}
  {\bibfield  {journal} {\bibinfo  {journal} {Rev. Mod. Phys.}\ }\textbf
  {\bibinfo {volume} {81}},\ \bibinfo {pages} {1301} (\bibinfo {year}
  {2009})}\BibitemShut {NoStop}%
\bibitem [{\citenamefont {Takahashi}\ \emph {et~al.}(2010)\citenamefont
  {Takahashi}, \citenamefont {Neergaard-Nielsen}, \citenamefont {Takeuchi},
  \citenamefont {Takeoka}, \citenamefont {Hayasaka}, \citenamefont {Furusawa},\
  and\ \citenamefont {Sasaki}}]{masahide.np.4.178.2010}%
  \BibitemOpen
  \bibfield  {author} {\bibinfo {author} {\bibfnamefont {H.}~\bibnamefont
  {Takahashi}}, \bibinfo {author} {\bibfnamefont {J.~S.}\ \bibnamefont
  {Neergaard-Nielsen}}, \bibinfo {author} {\bibfnamefont {M.}~\bibnamefont
  {Takeuchi}}, \bibinfo {author} {\bibfnamefont {M.}~\bibnamefont {Takeoka}},
  \bibinfo {author} {\bibfnamefont {K.}~\bibnamefont {Hayasaka}}, \bibinfo
  {author} {\bibfnamefont {A.}~\bibnamefont {Furusawa}}, \ and\ \bibinfo
  {author} {\bibfnamefont {M.}~\bibnamefont {Sasaki}},\ }\href
  {https://www.nature.com/articles/nphoton.2010.1} {\bibfield  {journal}
  {\bibinfo  {journal} {Nat. Photonics}\ }\textbf {\bibinfo {volume} {4}},\
  \bibinfo {pages} {178} (\bibinfo {year} {2010})}\BibitemShut {NoStop}%
\bibitem [{\citenamefont {Eisert}\ \emph {et~al.}(2002)\citenamefont {Eisert},
  \citenamefont {Scheel},\ and\ \citenamefont
  {Plenio}}]{eieret.prl.89.137903.2002}%
  \BibitemOpen
  \bibfield  {author} {\bibinfo {author} {\bibfnamefont {J.}~\bibnamefont
  {Eisert}}, \bibinfo {author} {\bibfnamefont {S.}~\bibnamefont {Scheel}}, \
  and\ \bibinfo {author} {\bibfnamefont {M.~B.}\ \bibnamefont {Plenio}},\
  }\href {\doibase 10.1103/PhysRevLett.89.137903} {\bibfield  {journal}
  {\bibinfo  {journal} {Phys. Rev. Lett.}\ }\textbf {\bibinfo {volume} {89}},\
  \bibinfo {pages} {137903} (\bibinfo {year} {2002})}\BibitemShut {NoStop}%
\bibitem [{\citenamefont {Lloyd}\ and\ \citenamefont
  {Braunstein}(1999)}]{l1oyd.prl.82.1784.1999}%
  \BibitemOpen
  \bibfield  {author} {\bibinfo {author} {\bibfnamefont {S.}~\bibnamefont
  {Lloyd}}\ and\ \bibinfo {author} {\bibfnamefont {S.~L.}\ \bibnamefont
  {Braunstein}},\ }\href {\doibase 10.1103/PhysRevLett.82.1784} {\bibfield
  {journal} {\bibinfo  {journal} {Phys. Rev. Lett.}\ }\textbf {\bibinfo
  {volume} {82}},\ \bibinfo {pages} {1784} (\bibinfo {year}
  {1999})}\BibitemShut {NoStop}%
\bibitem [{\citenamefont {Bartlett}\ \emph {et~al.}(2002)\citenamefont
  {Bartlett}, \citenamefont {Sanders}, \citenamefont {Braunstein},\ and\
  \citenamefont {Nemoto}}]{bartlett.prl.88.097904.2002}%
  \BibitemOpen
  \bibfield  {author} {\bibinfo {author} {\bibfnamefont {S.~D.}\ \bibnamefont
  {Bartlett}}, \bibinfo {author} {\bibfnamefont {B.~C.}\ \bibnamefont
  {Sanders}}, \bibinfo {author} {\bibfnamefont {S.~L.}\ \bibnamefont
  {Braunstein}}, \ and\ \bibinfo {author} {\bibfnamefont {K.}~\bibnamefont
  {Nemoto}},\ }\href {\doibase 10.1103/PhysRevLett.88.097904} {\bibfield
  {journal} {\bibinfo  {journal} {Phys. Rev. Lett.}\ }\textbf {\bibinfo
  {volume} {88}},\ \bibinfo {pages} {097904} (\bibinfo {year}
  {2002})}\BibitemShut {NoStop}%
\bibitem [{\citenamefont {Menicucci}\ \emph {et~al.}(2006)\citenamefont
  {Menicucci}, \citenamefont {van Loock}, \citenamefont {Gu}, \citenamefont
  {Weedbrook}, \citenamefont {Ralph},\ and\ \citenamefont
  {Nielsen}}]{nielsen.prl.97.110501.2006}%
  \BibitemOpen
  \bibfield  {author} {\bibinfo {author} {\bibfnamefont {N.~C.}\ \bibnamefont
  {Menicucci}}, \bibinfo {author} {\bibfnamefont {P.}~\bibnamefont {van
  Loock}}, \bibinfo {author} {\bibfnamefont {M.}~\bibnamefont {Gu}}, \bibinfo
  {author} {\bibfnamefont {C.}~\bibnamefont {Weedbrook}}, \bibinfo {author}
  {\bibfnamefont {T.~C.}\ \bibnamefont {Ralph}}, \ and\ \bibinfo {author}
  {\bibfnamefont {M.~A.}\ \bibnamefont {Nielsen}},\ }\href {\doibase
  10.1103/PhysRevLett.97.110501} {\bibfield  {journal} {\bibinfo  {journal}
  {Phys. Rev. Lett.}\ }\textbf {\bibinfo {volume} {97}},\ \bibinfo {pages}
  {110501} (\bibinfo {year} {2006})}\BibitemShut {NoStop}%
\bibitem [{\citenamefont {Ohliger}\ \emph {et~al.}(2010)\citenamefont
  {Ohliger}, \citenamefont {Kieling},\ and\ \citenamefont
  {Eisert}}]{eisert.pra.82.042336.2010}%
  \BibitemOpen
  \bibfield  {author} {\bibinfo {author} {\bibfnamefont {M.}~\bibnamefont
  {Ohliger}}, \bibinfo {author} {\bibfnamefont {K.}~\bibnamefont {Kieling}}, \
  and\ \bibinfo {author} {\bibfnamefont {J.}~\bibnamefont {Eisert}},\ }\href
  {\doibase 10.1103/PhysRevA.82.042336} {\bibfield  {journal} {\bibinfo
  {journal} {Phys. Rev. A}\ }\textbf {\bibinfo {volume} {82}},\ \bibinfo
  {pages} {042336} (\bibinfo {year} {2010})}\BibitemShut {NoStop}%
\bibitem [{\citenamefont {Ohliger}\ and\ \citenamefont
  {Eisert}(2012)}]{eisert.pra.85.062318.2012}%
  \BibitemOpen
  \bibfield  {author} {\bibinfo {author} {\bibfnamefont {M.}~\bibnamefont
  {Ohliger}}\ and\ \bibinfo {author} {\bibfnamefont {J.}~\bibnamefont
  {Eisert}},\ }\href {\doibase 10.1103/PhysRevA.85.062318} {\bibfield
  {journal} {\bibinfo  {journal} {Phys. Rev. A}\ }\textbf {\bibinfo {volume}
  {85}},\ \bibinfo {pages} {062318} (\bibinfo {year} {2012})}\BibitemShut
  {NoStop}%
\bibitem [{\citenamefont {Ourjoumtsev}\ \emph {et~al.}(2007)\citenamefont
  {Ourjoumtsev}, \citenamefont {Dantan}, \citenamefont {Tualle-Brouri},\ and\
  \citenamefont {Grangier}}]{ourjoumtsev.prl.98.030502.2007}%
  \BibitemOpen
  \bibfield  {author} {\bibinfo {author} {\bibfnamefont {A.}~\bibnamefont
  {Ourjoumtsev}}, \bibinfo {author} {\bibfnamefont {A.}~\bibnamefont {Dantan}},
  \bibinfo {author} {\bibfnamefont {R.}~\bibnamefont {Tualle-Brouri}}, \ and\
  \bibinfo {author} {\bibfnamefont {P.}~\bibnamefont {Grangier}},\ }\href
  {\doibase 10.1103/PhysRevLett.98.030502} {\bibfield  {journal} {\bibinfo
  {journal} {Phys. Rev. Lett.}\ }\textbf {\bibinfo {volume} {98}},\ \bibinfo
  {pages} {030502} (\bibinfo {year} {2007})}\BibitemShut {NoStop}%
\bibitem [{\citenamefont {Ourjoumtsev}\ \emph {et~al.}(2009)\citenamefont
  {Ourjoumtsev}, \citenamefont {Ferreyrol}, \citenamefont {Tualle-Brouri},\
  and\ \citenamefont {Grangier}}]{ourjoumtsev.np.5.189.2009}%
  \BibitemOpen
  \bibfield  {author} {\bibinfo {author} {\bibfnamefont {A.}~\bibnamefont
  {Ourjoumtsev}}, \bibinfo {author} {\bibfnamefont {F.}~\bibnamefont
  {Ferreyrol}}, \bibinfo {author} {\bibfnamefont {R.}~\bibnamefont
  {Tualle-Brouri}}, \ and\ \bibinfo {author} {\bibfnamefont {P.}~\bibnamefont
  {Grangier}},\ }\href {https://www.nature.com/articles/nphys1199} {\bibfield
  {journal} {\bibinfo  {journal} {Nat. Phys.}\ }\textbf {\bibinfo {volume}
  {5}},\ \bibinfo {pages} {189} (\bibinfo {year} {2009})}\BibitemShut {NoStop}%
\bibitem [{\citenamefont {Ourjoumtsev}\ \emph {et~al.}(2006)\citenamefont
  {Ourjoumtsev}, \citenamefont {Tualle-Brouri}, \citenamefont {Laurat},\ and\
  \citenamefont {Grangier}}]{ourjoumtsev.science.312.83.2006}%
  \BibitemOpen
  \bibfield  {author} {\bibinfo {author} {\bibfnamefont {A.}~\bibnamefont
  {Ourjoumtsev}}, \bibinfo {author} {\bibfnamefont {R.}~\bibnamefont
  {Tualle-Brouri}}, \bibinfo {author} {\bibfnamefont {J.}~\bibnamefont
  {Laurat}}, \ and\ \bibinfo {author} {\bibfnamefont {P.}~\bibnamefont
  {Grangier}},\ }\href {https://science.sciencemag.org/content/312/5770/83}
  {\bibfield  {journal} {\bibinfo  {journal} {Science}\ }\textbf {\bibinfo
  {volume} {312}},\ \bibinfo {pages} {83} (\bibinfo {year} {2006})}\BibitemShut
  {NoStop}%
\bibitem [{\citenamefont {Sychev}\ \emph {et~al.}(2017)\citenamefont {Sychev},
  \citenamefont {Ulanov}, \citenamefont {Pushkina}, \citenamefont {Richards},
  \citenamefont {Fedorov},\ and\ \citenamefont
  {Lvovsky}}]{sychev.np.11.379.2017}%
  \BibitemOpen
  \bibfield  {author} {\bibinfo {author} {\bibfnamefont {D.~V.}\ \bibnamefont
  {Sychev}}, \bibinfo {author} {\bibfnamefont {A.~E.}\ \bibnamefont {Ulanov}},
  \bibinfo {author} {\bibfnamefont {A.~A.}\ \bibnamefont {Pushkina}}, \bibinfo
  {author} {\bibfnamefont {M.~W.}\ \bibnamefont {Richards}}, \bibinfo {author}
  {\bibfnamefont {I.~A.}\ \bibnamefont {Fedorov}}, \ and\ \bibinfo {author}
  {\bibfnamefont {A.~I.}\ \bibnamefont {Lvovsky}},\ }\href
  {https://www.nature.com/articles/nphoton.2017.57} {\bibfield  {journal}
  {\bibinfo  {journal} {Nat. Photonics}\ }\textbf {\bibinfo {volume} {11}},\
  \bibinfo {pages} {379} (\bibinfo {year} {2017})}\BibitemShut {NoStop}%
\bibitem [{\citenamefont {Serikawa}\ \emph {et~al.}(2018)\citenamefont
  {Serikawa}, \citenamefont {Yoshikawa}, \citenamefont {Takeda}, \citenamefont
  {Yonezawa}, \citenamefont {Ralph}, \citenamefont {Huntington},\ and\
  \citenamefont {Furusawa}}]{furusawa.prl.121.143602.2018}%
  \BibitemOpen
  \bibfield  {author} {\bibinfo {author} {\bibfnamefont {T.}~\bibnamefont
  {Serikawa}}, \bibinfo {author} {\bibfnamefont {J.-i.}\ \bibnamefont
  {Yoshikawa}}, \bibinfo {author} {\bibfnamefont {S.}~\bibnamefont {Takeda}},
  \bibinfo {author} {\bibfnamefont {H.}~\bibnamefont {Yonezawa}}, \bibinfo
  {author} {\bibfnamefont {T.~C.}\ \bibnamefont {Ralph}}, \bibinfo {author}
  {\bibfnamefont {E.~H.}\ \bibnamefont {Huntington}}, \ and\ \bibinfo {author}
  {\bibfnamefont {A.}~\bibnamefont {Furusawa}},\ }\href {\doibase
  10.1103/PhysRevLett.121.143602} {\bibfield  {journal} {\bibinfo  {journal}
  {Phys. Rev. Lett.}\ }\textbf {\bibinfo {volume} {121}},\ \bibinfo {pages}
  {143602} (\bibinfo {year} {2018})}\BibitemShut {NoStop}%
\bibitem [{\citenamefont {Biagi}\ \emph {et~al.}(2020)\citenamefont {Biagi},
  \citenamefont {Costanzo}, \citenamefont {Bellini},\ and\ \citenamefont
  {Zavatta}}]{nicola.prl.124.033604.2020}%
  \BibitemOpen
  \bibfield  {author} {\bibinfo {author} {\bibfnamefont {N.}~\bibnamefont
  {Biagi}}, \bibinfo {author} {\bibfnamefont {L.~S.}\ \bibnamefont {Costanzo}},
  \bibinfo {author} {\bibfnamefont {M.}~\bibnamefont {Bellini}}, \ and\
  \bibinfo {author} {\bibfnamefont {A.}~\bibnamefont {Zavatta}},\ }\href
  {\doibase 10.1103/PhysRevLett.124.033604} {\bibfield  {journal} {\bibinfo
  {journal} {Phys. Rev. Lett.}\ }\textbf {\bibinfo {volume} {124}},\ \bibinfo
  {pages} {033604} (\bibinfo {year} {2020})}\BibitemShut {NoStop}%
\bibitem [{\citenamefont {Jeong}\ \emph {et~al.}(2014)\citenamefont {Jeong},
  \citenamefont {Zavatta}, \citenamefont {Kang}, \citenamefont {Lee},
  \citenamefont {Costanzo}, \citenamefont {Grandi}, \citenamefont {Ralph},\
  and\ \citenamefont {Bellini}}]{jeong.np.8.564.2014}%
  \BibitemOpen
  \bibfield  {author} {\bibinfo {author} {\bibfnamefont {H.}~\bibnamefont
  {Jeong}}, \bibinfo {author} {\bibfnamefont {A.}~\bibnamefont {Zavatta}},
  \bibinfo {author} {\bibfnamefont {M.}~\bibnamefont {Kang}}, \bibinfo {author}
  {\bibfnamefont {S.-W.}\ \bibnamefont {Lee}}, \bibinfo {author} {\bibfnamefont
  {L.~S.}\ \bibnamefont {Costanzo}}, \bibinfo {author} {\bibfnamefont
  {S.}~\bibnamefont {Grandi}}, \bibinfo {author} {\bibfnamefont {T.~C.}\
  \bibnamefont {Ralph}}, \ and\ \bibinfo {author} {\bibfnamefont
  {M.}~\bibnamefont {Bellini}},\ }\href
  {https://www.nature.com/articles/nphoton.2014.136} {\bibfield  {journal}
  {\bibinfo  {journal} {Nat. Photonics}\ }\textbf {\bibinfo {volume} {8}},\
  \bibinfo {pages} {564} (\bibinfo {year} {2014})}\BibitemShut {NoStop}%
\bibitem [{\citenamefont {Morin}\ \emph {et~al.}(2014)\citenamefont {Morin},
  \citenamefont {Huang}, \citenamefont {Liu}, \citenamefont {Le~Jeannic},
  \citenamefont {Fabre},\ and\ \citenamefont {Laurat}}]{morin.np.8.570.2014}%
  \BibitemOpen
  \bibfield  {author} {\bibinfo {author} {\bibfnamefont {O.}~\bibnamefont
  {Morin}}, \bibinfo {author} {\bibfnamefont {K.}~\bibnamefont {Huang}},
  \bibinfo {author} {\bibfnamefont {J.}~\bibnamefont {Liu}}, \bibinfo {author}
  {\bibfnamefont {H.}~\bibnamefont {Le~Jeannic}}, \bibinfo {author}
  {\bibfnamefont {C.}~\bibnamefont {Fabre}}, \ and\ \bibinfo {author}
  {\bibfnamefont {J.}~\bibnamefont {Laurat}},\ }\href
  {https://www.nature.com/articles/nphoton.2014.137} {\bibfield  {journal}
  {\bibinfo  {journal} {Nat. Photonics}\ }\textbf {\bibinfo {volume} {8}},\
  \bibinfo {pages} {570} (\bibinfo {year} {2014})}\BibitemShut {NoStop}%
\bibitem [{\citenamefont {Ra}\ \emph {et~al.}(2019)\citenamefont {Ra},
  \citenamefont {Dufour}, \citenamefont {Walschaers}, \citenamefont {Jacquard},
  \citenamefont {Michel}, \citenamefont {Fabre},\ and\ \citenamefont
  {Treps}}]{ra.np.2019non}%
  \BibitemOpen
  \bibfield  {author} {\bibinfo {author} {\bibfnamefont {Y.-S.}\ \bibnamefont
  {Ra}}, \bibinfo {author} {\bibfnamefont {A.}~\bibnamefont {Dufour}}, \bibinfo
  {author} {\bibfnamefont {M.}~\bibnamefont {Walschaers}}, \bibinfo {author}
  {\bibfnamefont {C.}~\bibnamefont {Jacquard}}, \bibinfo {author}
  {\bibfnamefont {T.}~\bibnamefont {Michel}}, \bibinfo {author} {\bibfnamefont
  {C.}~\bibnamefont {Fabre}}, \ and\ \bibinfo {author} {\bibfnamefont
  {N.}~\bibnamefont {Treps}},\ }\href
  {https://www.nature.com/articles/s41567-019-0726-y} {\bibfield  {journal}
  {\bibinfo  {journal} {Nat. Phys.}\ ,\ \bibinfo {pages} {1}} (\bibinfo {year}
  {2019})}\BibitemShut {NoStop}%
\bibitem [{\citenamefont {Douady}\ and\ \citenamefont
  {Boulanger}(2004)}]{douady.23.2794.ol.2004}%
  \BibitemOpen
  \bibfield  {author} {\bibinfo {author} {\bibfnamefont {J.}~\bibnamefont
  {Douady}}\ and\ \bibinfo {author} {\bibfnamefont {B.}~\bibnamefont
  {Boulanger}},\ }\href {\doibase 10.1364/OL.29.002794} {\bibfield  {journal}
  {\bibinfo  {journal} {Opt. Lett.}\ }\textbf {\bibinfo {volume} {29}},\
  \bibinfo {pages} {2794} (\bibinfo {year} {2004})}\BibitemShut {NoStop}%
\bibitem [{\citenamefont {Chang}\ \emph {et~al.}(2020)\citenamefont {Chang},
  \citenamefont {Sab\'{\i}n}, \citenamefont {Forn-D\'{\i}az}, \citenamefont
  {Quijandr\'{\i}a}, \citenamefont {Vadiraj}, \citenamefont {Nsanzineza},
  \citenamefont {Johansson},\ and\ \citenamefont
  {Wilson}}]{chang.prx.10.011011.2020}%
  \BibitemOpen
  \bibfield  {author} {\bibinfo {author} {\bibfnamefont {C.~W.~S.}\
  \bibnamefont {Chang}}, \bibinfo {author} {\bibfnamefont {C.}~\bibnamefont
  {Sab\'{\i}n}}, \bibinfo {author} {\bibfnamefont {P.}~\bibnamefont
  {Forn-D\'{\i}az}}, \bibinfo {author} {\bibfnamefont {F.}~\bibnamefont
  {Quijandr\'{\i}a}}, \bibinfo {author} {\bibfnamefont {A.~M.}\ \bibnamefont
  {Vadiraj}}, \bibinfo {author} {\bibfnamefont {I.}~\bibnamefont {Nsanzineza}},
  \bibinfo {author} {\bibfnamefont {G.}~\bibnamefont {Johansson}}, \ and\
  \bibinfo {author} {\bibfnamefont {C.~M.}\ \bibnamefont {Wilson}},\ }\href
  {\doibase 10.1103/PhysRevX.10.011011} {\bibfield  {journal} {\bibinfo
  {journal} {Phys. Rev. X}\ }\textbf {\bibinfo {volume} {10}},\ \bibinfo
  {pages} {011011} (\bibinfo {year} {2020})}\BibitemShut {NoStop}%
\bibitem [{\citenamefont {Felbinger}\ \emph {et~al.}(1998)\citenamefont
  {Felbinger}, \citenamefont {Schiller},\ and\ \citenamefont
  {Mlynek}}]{felbinger.prl.80.492.1998}%
  \BibitemOpen
  \bibfield  {author} {\bibinfo {author} {\bibfnamefont {T.}~\bibnamefont
  {Felbinger}}, \bibinfo {author} {\bibfnamefont {S.}~\bibnamefont {Schiller}},
  \ and\ \bibinfo {author} {\bibfnamefont {J.}~\bibnamefont {Mlynek}},\ }\href
  {\doibase 10.1103/PhysRevLett.80.492} {\bibfield  {journal} {\bibinfo
  {journal} {Phys. Rev. Lett.}\ }\textbf {\bibinfo {volume} {80}},\ \bibinfo
  {pages} {492} (\bibinfo {year} {1998})}\BibitemShut {NoStop}%
\bibitem [{\citenamefont {Bencheikh}\ \emph {et~al.}(2007)\citenamefont
  {Bencheikh}, \citenamefont {Gravier}, \citenamefont {Douady}, \citenamefont
  {Levenson},\ and\ \citenamefont {Boulanger}}]{kamel.crp.8.206.2007}%
  \BibitemOpen
  \bibfield  {author} {\bibinfo {author} {\bibfnamefont {K.}~\bibnamefont
  {Bencheikh}}, \bibinfo {author} {\bibfnamefont {F.}~\bibnamefont {Gravier}},
  \bibinfo {author} {\bibfnamefont {J.}~\bibnamefont {Douady}}, \bibinfo
  {author} {\bibfnamefont {A.}~\bibnamefont {Levenson}}, \ and\ \bibinfo
  {author} {\bibfnamefont {B.}~\bibnamefont {Boulanger}},\ }\href {\doibase
  https://doi.org/10.1016/j.crhy.2006.07.014} {\bibfield  {journal} {\bibinfo
  {journal} {C. R. Phys.}\ }\textbf {\bibinfo {volume} {8}},\ \bibinfo {pages}
  {206 } (\bibinfo {year} {2007})}\BibitemShut {NoStop}%
\bibitem [{\citenamefont {Shen}\ \emph {et~al.}(2015)\citenamefont {Shen},
  \citenamefont {Assad}, \citenamefont {Grosse}, \citenamefont {Li},
  \citenamefont {Reid},\ and\ \citenamefont {Lam}}]{lam.prl.114.100403.2015}%
  \BibitemOpen
  \bibfield  {author} {\bibinfo {author} {\bibfnamefont {Y.}~\bibnamefont
  {Shen}}, \bibinfo {author} {\bibfnamefont {S.~M.}\ \bibnamefont {Assad}},
  \bibinfo {author} {\bibfnamefont {N.~B.}\ \bibnamefont {Grosse}}, \bibinfo
  {author} {\bibfnamefont {X.~Y.}\ \bibnamefont {Li}}, \bibinfo {author}
  {\bibfnamefont {M.~D.}\ \bibnamefont {Reid}}, \ and\ \bibinfo {author}
  {\bibfnamefont {P.~K.}\ \bibnamefont {Lam}},\ }\href {\doibase
  10.1103/PhysRevLett.114.100403} {\bibfield  {journal} {\bibinfo  {journal}
  {Phys. Rev. Lett.}\ }\textbf {\bibinfo {volume} {114}},\ \bibinfo {pages}
  {100403} (\bibinfo {year} {2015})}\BibitemShut {NoStop}%
\bibitem [{\citenamefont {Gonz\'alez}\ \emph {et~al.}(2018)\citenamefont
  {Gonz\'alez}, \citenamefont {Borne}, \citenamefont {Boulanger}, \citenamefont
  {Levenson},\ and\ \citenamefont {Bencheikh}}]{kamel.prl.120.043601.2018}%
  \BibitemOpen
  \bibfield  {author} {\bibinfo {author} {\bibfnamefont {E.~A.~R.}\
  \bibnamefont {Gonz\'alez}}, \bibinfo {author} {\bibfnamefont
  {A.}~\bibnamefont {Borne}}, \bibinfo {author} {\bibfnamefont
  {B.}~\bibnamefont {Boulanger}}, \bibinfo {author} {\bibfnamefont {J.~A.}\
  \bibnamefont {Levenson}}, \ and\ \bibinfo {author} {\bibfnamefont
  {K.}~\bibnamefont {Bencheikh}},\ }\href {\doibase
  10.1103/PhysRevLett.120.043601} {\bibfield  {journal} {\bibinfo  {journal}
  {Phys. Rev. Lett.}\ }\textbf {\bibinfo {volume} {120}},\ \bibinfo {pages}
  {043601} (\bibinfo {year} {2018})}\BibitemShut {NoStop}%
\bibitem [{\citenamefont {Dalibard}\ \emph {et~al.}(1992)\citenamefont
  {Dalibard}, \citenamefont {Castin},\ and\ \citenamefont
  {M\o{}lmer}}]{molmer.prl.68.580.1993}%
  \BibitemOpen
  \bibfield  {author} {\bibinfo {author} {\bibfnamefont {J.}~\bibnamefont
  {Dalibard}}, \bibinfo {author} {\bibfnamefont {Y.}~\bibnamefont {Castin}}, \
  and\ \bibinfo {author} {\bibfnamefont {K.}~\bibnamefont {M\o{}lmer}},\ }\href
  {\doibase 10.1103/PhysRevLett.68.580} {\bibfield  {journal} {\bibinfo
  {journal} {Phys. Rev. Lett.}\ }\textbf {\bibinfo {volume} {68}},\ \bibinfo
  {pages} {580} (\bibinfo {year} {1992})}\BibitemShut {NoStop}%
\bibitem [{\citenamefont {M{\o}lmer}\ \emph {et~al.}(1993)\citenamefont
  {M{\o}lmer}, \citenamefont {Castin},\ and\ \citenamefont
  {Dalibard}}]{molmer.josab.10.524.1993}%
  \BibitemOpen
  \bibfield  {author} {\bibinfo {author} {\bibfnamefont {K.}~\bibnamefont
  {M{\o}lmer}}, \bibinfo {author} {\bibfnamefont {Y.}~\bibnamefont {Castin}}, \
  and\ \bibinfo {author} {\bibfnamefont {J.}~\bibnamefont {Dalibard}},\ }\href
  {\doibase 10.1364/JOSAB.10.000524} {\bibfield  {journal} {\bibinfo  {journal}
  {J. Opt. Soc. Am. B}\ }\textbf {\bibinfo {volume} {10}},\ \bibinfo {pages}
  {524} (\bibinfo {year} {1993})}\BibitemShut {NoStop}%
\bibitem [{\citenamefont {Johansson}\ \emph {et~al.}(2013)\citenamefont
  {Johansson}, \citenamefont {Nation},\ and\ \citenamefont {Nori}}]{qutip1}%
  \BibitemOpen
  \bibfield  {author} {\bibinfo {author} {\bibfnamefont {J.}~\bibnamefont
  {Johansson}}, \bibinfo {author} {\bibfnamefont {P.}~\bibnamefont {Nation}}, \
  and\ \bibinfo {author} {\bibfnamefont {F.}~\bibnamefont {Nori}},\ }\href
  {\doibase https://doi.org/10.1016/j.cpc.2012.11.019} {\bibfield  {journal}
  {\bibinfo  {journal} {Comp. Phys. Comm.}\ }\textbf {\bibinfo {volume}
  {184}},\ \bibinfo {pages} {1234 } (\bibinfo {year} {2013})}\BibitemShut
  {NoStop}%
\bibitem [{\citenamefont {Johansson}\ \emph {et~al.}(2012)\citenamefont
  {Johansson}, \citenamefont {Nation},\ and\ \citenamefont {Nori}}]{qutip2}%
  \BibitemOpen
  \bibfield  {author} {\bibinfo {author} {\bibfnamefont {J.}~\bibnamefont
  {Johansson}}, \bibinfo {author} {\bibfnamefont {P.}~\bibnamefont {Nation}}, \
  and\ \bibinfo {author} {\bibfnamefont {F.}~\bibnamefont {Nori}},\ }\href
  {\doibase https://doi.org/10.1016/j.cpc.2012.02.021} {\bibfield  {journal}
  {\bibinfo  {journal} {Comp. Phys. Comm.}\ }\textbf {\bibinfo {volume}
  {183}},\ \bibinfo {pages} {1760 } (\bibinfo {year} {2012})}\BibitemShut
  {NoStop}%
\bibitem [{\citenamefont {Kok}\ and\ \citenamefont
  {Lovett}(2010)}]{kok.quantuminformationprocessing.2010}%
  \BibitemOpen
  \bibfield  {author} {\bibinfo {author} {\bibfnamefont {P.}~\bibnamefont
  {Kok}}\ and\ \bibinfo {author} {\bibfnamefont {B.~W.}\ \bibnamefont
  {Lovett}},\ }\href@noop {} {\emph {\bibinfo {title} {Introduction to optical
  quantum information processing}}}\ (\bibinfo  {publisher} {Cambridge
  university press},\ \bibinfo {year} {2010})\BibitemShut {NoStop}%
\bibitem [{\citenamefont {Agarwal}(2012)}]{agarwal2012quantum}%
  \BibitemOpen
  \bibfield  {author} {\bibinfo {author} {\bibfnamefont {G.~S.}\ \bibnamefont
  {Agarwal}},\ }\href@noop {} {\emph {\bibinfo {title} {Quantum optics}}}\
  (\bibinfo  {publisher} {Cambridge University Press},\ \bibinfo {year}
  {2012})\BibitemShut {NoStop}%
\bibitem [{\citenamefont {Genoni}\ \emph {et~al.}(2008)\citenamefont {Genoni},
  \citenamefont {Paris},\ and\ \citenamefont {Banaszek}}]{marco.pra.78.6.2008}%
  \BibitemOpen
  \bibfield  {author} {\bibinfo {author} {\bibfnamefont {M.~G.}\ \bibnamefont
  {Genoni}}, \bibinfo {author} {\bibfnamefont {M.~G.~A.}\ \bibnamefont
  {Paris}}, \ and\ \bibinfo {author} {\bibfnamefont {K.}~\bibnamefont
  {Banaszek}},\ }\href {\doibase 10.1103/PhysRevA.78.060303} {\bibfield
  {journal} {\bibinfo  {journal} {Phys. Rev. A}\ }\textbf {\bibinfo {volume}
  {78}},\ \bibinfo {pages} {060303} (\bibinfo {year} {2008})}\BibitemShut
  {NoStop}%
\bibitem [{\citenamefont {Vidal}\ and\ \citenamefont
  {Werner}(2002)}]{vidal.pra.65.032314.2002}%
  \BibitemOpen
  \bibfield  {author} {\bibinfo {author} {\bibfnamefont {G.}~\bibnamefont
  {Vidal}}\ and\ \bibinfo {author} {\bibfnamefont {R.~F.}\ \bibnamefont
  {Werner}},\ }\href {\doibase 10.1103/PhysRevA.65.032314} {\bibfield
  {journal} {\bibinfo  {journal} {Phys. Rev. A}\ }\textbf {\bibinfo {volume}
  {65}},\ \bibinfo {pages} {032314} (\bibinfo {year} {2002})}\BibitemShut
  {NoStop}%
\bibitem [{\citenamefont {Simon}(2000)}]{simon.prl.84.2726.2000}%
  \BibitemOpen
  \bibfield  {author} {\bibinfo {author} {\bibfnamefont {R.}~\bibnamefont
  {Simon}},\ }\href {\doibase 10.1103/PhysRevLett.84.2726} {\bibfield
  {journal} {\bibinfo  {journal} {Phys. Rev. Lett.}\ }\textbf {\bibinfo
  {volume} {84}},\ \bibinfo {pages} {2726} (\bibinfo {year}
  {2000})}\BibitemShut {NoStop}%
\bibitem [{\citenamefont {at~http://link.aps.org/supplemental/XXXXXX for
  details on the calculation of~the Quantum Relative~Entropy}\ and\
  \citenamefont {the~optical losses.}()}]{sm.prl.2020}%
  \BibitemOpen
  \bibfield  {author} {\bibinfo {author} {\bibfnamefont {}\
  \bibnamefont {See Supplemental Material at http://link.aps.org/~supplemental/XXXXXX for details on the
  calculation of~the quantum relative~entropy}}\ and\ \bibinfo {author}
  {\bibnamefont {the~optical losses.}}\ }
\bibitem [{\citenamefont {Armstrong}\ \emph {et~al.}(2015)\citenamefont
  {Armstrong}, \citenamefont {Wang}, \citenamefont {Teh}, \citenamefont {Gong},
  \citenamefont {He}, \citenamefont {Janousek}, \citenamefont {Bachor},
  \citenamefont {Reid},\ and\ \citenamefont {Lam}}]{armstrong.np.11.2015}%
  \BibitemOpen
  \bibfield  {author} {\bibinfo {author} {\bibfnamefont {S.}~\bibnamefont
  {Armstrong}}, \bibinfo {author} {\bibfnamefont {M.}~\bibnamefont {Wang}},
  \bibinfo {author} {\bibfnamefont {R.~Y.}\ \bibnamefont {Teh}}, \bibinfo
  {author} {\bibfnamefont {Q.}~\bibnamefont {Gong}}, \bibinfo {author}
  {\bibfnamefont {Q.}~\bibnamefont {He}}, \bibinfo {author} {\bibfnamefont
  {J.}~\bibnamefont {Janousek}}, \bibinfo {author} {\bibfnamefont {H.-A.}\
  \bibnamefont {Bachor}}, \bibinfo {author} {\bibfnamefont {M.~D.}\
  \bibnamefont {Reid}}, \ and\ \bibinfo {author} {\bibfnamefont {P.~K.}\
  \bibnamefont {Lam}},\ }\href
  {https://www.nature.com/articles/nphys3202#citeas} {\bibfield  {journal}
  {\bibinfo  {journal} {Nat. Phys.}\ }\textbf {\bibinfo {volume} {11}},\
  \bibinfo {pages} {167} (\bibinfo {year} {2015})}\BibitemShut {NoStop}%
\end{thebibliography}

%

\end{document}